\documentclass{iopart}
\eqnobysec
\usepackage{iopams} % AMS fonts and some macros a la IOP
\usepackage{graphicx}
\usepackage{hyperref}
\newcommand{\mathid}[1]{\mathrm{#1}}
\newcommand{\mathvec}[1]{\boldsymbol{#1}}

\begin{document}
\title[Non-adiabatic effects during the dissociative adsorption of O$_{2}$ at Ag(111)?]
{Non-adiabatic effects during the dissociative adsorption of O$_{2}$ at Ag(111)?\\
A first-principles divide and conquer study}
\author{Itziar Goikoetxea$^1$, Juan Beltr{\'a}n$^2$, J{\"o}rg Meyer$^{2,3}$,
J. I{\~n}aki Juaristi$^{4,1,5}$, Maite Alducin$^{1,5}$, and Karsten Reuter$^{2,3}$}
\address{$^1$ Centro de F\'{\i}sica de Materiales CFM/MPC (CSIC-UPV/EHU), San Sebasti{\'a}n, Spain}
\address{$^2$ Fritz-Haber-Institut der Max-Planck-Gesellschaft, Faradayweg 4-6, D-14195 Berlin, Germany}
\address{$^3$ Department Chemie, Technische Universit{\"a}t M{\"u}nchen, Lichtenbergstr. 4, D-85747 Garching, Germany}
\address{$^4$ Departamento de F\'{\i}sica de Materiales, Facultad de Qu\'{\i}micas, UPV/EHU, San Sebasti{\'a}n, Spain}
\address{$^5$ Donostia International Physics Center DIPC, San Sebasti{\'a}n, Spain}
\ead{itziar\_goicoechea@ehu.es}

\begin{abstract}
We study the gas-surface dynamics of O$_2$ at Ag(111) with the
particular objective to unravel whether electronic non-adiabatic
effects are contributing to the experimentally established
inertness of the surface with respect to oxygen uptake. We
employ a first-principles divide and conquer approach based on
an extensive density-functional theory mapping of the adiabatic
potential energy surface (PES) along the six O$_2$ molecular
degrees of freedom. Neural networks are subsequently used to
interpolate this grid data to a continuous representation. The
low computational cost with which forces are available from
this PES representation allows then for a sufficiently large
number of molecular dynamics trajectories to quantitatively
determine the very low initial dissociative sticking coefficient
at this surface. Already these adiabatic calculations yield
dissociation probabilities close to the scattered experimental
data. Our analysis shows that this low reactivity is governed
by large energy barriers in excess of 1.1~eV very close to the
surface. Unfortunately, these adiabatic PES characteristics
render the dissociative sticking a rather insensitive quantity
with respect to a potential spin or charge non-adiabaticity in
the O$_2$-Ag(111) interaction. We correspondingly attribute
the remaining deviations between the computed and measured
dissociation probabilities primarily to unresolved experimental
issues with respect to surface imperfections.
\end{abstract}

% 34.35.+a  Interactions of atoms and molecules with surface
% 79.20.Ap  Theory of impact phenomena; numerical simulation
% 79.20.Rf  Atomic, molecular, and ion beam impact and interactions with surfaces

\submitto{\NJP}

\maketitle

\section{Introduction\label{intro}}

A predictive materials science modelling based on microscopic
understanding requires a thorough knowledge of all underlying
elementary processes at the atomic scale. The (dissociative)
adsorption of individual oxygen molecules at metal surfaces is
such an elementary process that is of crucial relevance for
a wide range of applications, with heterogeneous catalysis
forming just one prominent example. Recent work on the
dissociative adsorption of O$_{2}$ at Al(111) has severely
challenged the prevalent understanding of this process
\cite{behler05,behler08,carbogno08,carbogno10}.
Significant spin non-adiabatic effects, i.e. a coupled
electronic-nuclear motion beyond the standard Born-Oppenheimer
approximation, have been proposed as rationalization for the
experimentally observed low initial adsorption probability. This
behaviour has been associated to the low density-of-states
(DOS) at the Fermi-level of the simple metal surface, which
prevents an efficient spin quenching through the tunnelling
of electrons between substrate and adsorbate. This hinders the
transition from the spin-triplet ground state of gas-phase O$_2$
into the singlet state upon adsorption at the metal surface.
The low sticking coefficient can then be seen as a result
of similar spin-selection rules that are well known from
gas-phase chemistry.

In particular from the point of view of heterogeneous
catalysis, interest naturally shifts to low-index noble metal
surfaces. Only recently, a spin-asymmetry in electron-hole pair
excitation spectra for O$_{2}$ at Pd(100) have further fortified
the previously proposed DOS-related mechanism~\cite{meyer11}. On
the other hand, in contrast to palladium, the DOS at the
Fermi-level is equally low for aluminium and coinage
metals. This raises the question whether similar spin
non-adiabatic effects significantly contribute to their
established low reactivity with respect to O$_2$ dissociation
\cite{vattuone94jcp,vattuone94prl,raukema96jpcm,vattuone98,raukema96ss,chesters75,canning84,deng05}.
Here, we aim to contribute to this with a first-principles based
modelling of the O$_2$ gas-surface dynamics specifically
at the Ag(111) surface, for which particularly low
adsorption probabilities have been reported experimentally
\cite{raukema96ss,rocca97}. Earlier simplified models based
on low dimensional analytic potentials have already tried to
understand the nature of such low reactivity in terms of charge
transfer limitations from the Ag(111) surface to the impinging
O$_2$ molecule \cite{citri96,zhdanov97}. Notwithstanding, on
Ag(100) recent molecular dynamics (MD) calculations based on an
{\em ab initio} six-dimensional potential energy surface (PES)
showed that the comparable absence of O$_2$ dissociation on
that surface can be fully explained within an adiabatic picture
\cite{alducin08}. There, the system is simply characterized
by large dissociation barriers of about 1.1\,eV that appear
when the molecule is close to the surface. This motivates a
similar {\em ab initio} investigation also for Ag(111), with
the objective to settle the question up to what extent the
low reactivity on the Ag(111) surface is really a signature
of electronic non-adiabaticity or only the result of large
dissociation barriers as found for Ag(100).

The overall structure of the paper is as follows. The next
section provides a detailed description of the different steps
employed to model the dynamics of O$_2$ on Ag(111) within a
divide-and-conquer approach. \Sref{s:results} commences with
a brief review of the existing experimental data, emphasizing
controversies with respect to the absolute order of magnitude
of dissociative sticking for incidence energies $E_i$ below
1\,eV. We then proceed with a discussion of the quality and
relevant static properties of the computed adiabatic PES,
revealing as a central characteristic barriers to dissociation
larger than 1.1\,eV at close distances to the surface. The
ensuing classical trajectory calculations demonstrate
that these barriers lead to dissociation probabilities
that already fall within the large error bars set by the
scattered experimental data. With heights that are larger
than the gas-phase O$_2$ spin singlet-triplet gap (0.98\,eV
\cite{bookami}) and the O$_2$ electron affinity (0.44\,eV
\cite{bookami}) these barriers furthermore completely mask
any signs of a potential spin or charge non-adiabaticity in
the O$_2$-Ag(111) interaction in the dissociative sticking
coefficient. The remaining deviations of the computed sticking
curve to the measurements for $E_i < 1$\,eV are thus more likely
the result of the unresolved issues in the experimental data,
e.g. with respect to surface imperfections as discussed by
the experimentalists \cite{rocca97,kleyn96}.

\section{Methodology}

Methodologically, the very low dissociation that characterizes
this system makes the use of on-the-fly {\em ab initio}
molecular dynamics (AIMD) to determine the sticking
coefficient \cite{gross10} computationally unaffordable.
Instead, our simulations are based on a divide-and-conquer
approach \cite{engdahl92,gross95,wiesenekker96}, in which
an accurate PES is first constructed from first-principles,
and next, classical MD calculations are performed using
a continuous representation of this PES to describe the
molecule-surface interaction. Once the former is obtained,
the latter come at negligible computational cost. In contrast
to AIMD, the divide-and-conquer approach thus enables the
calculation of a large number of trajectories. Even for very
small sticking probabilities and various initial conditions of
molecules in the impinging molecular beam ($E_i$, incidence
angle etc.), sufficient statistical sampling can thus be
performed. Additionally, detailed information on the PES
properties, which allows for a better rationalization of
the factors ruling the gas-surface dynamics, provide another
advantage. Simulations based on this methodology have been able
to reproduce not only reactive probabilities, but also more
subtle quantities like the rovibrational state distributions
of scattered molecules \cite{diaz09,geethalakshmi11}.

The key ingredient of this impressive success that
ensures reliable quantitative results is the use of
multi-dimensional {\em ab initio}-based PESs. In this
respect, several interpolation methods to obtain continuous
representations in the context of gas-surface dynamics have
been developed in the last years, reaching energy precisions
on the order of meV in parts most relevant for the former
\cite{busnengo00,ischtwan94,lorenz04,lorenz06,behler07}. In
the present work, the PES interpolation is based on
neural networks (NN) \cite{lorenz04,lorenz06,behler07}
that we adapt to the specific symmetry of the O$_2$/Ag(111)
system as explained in \sref{s:nn}. In order to keep the
multi-dimensionality of the O$_2$/Ag(111) PES in a tractable
limit, we follow common practice and adopt the frozen surface
approximation.  This approximation has been successfully
applied to simulate the reactivity of various diatomic
molecules (H$_2$, N$_2$, O$_2$, \ldots) on metal surfaces
\cite{behler05,diaz09,pijper01,alducin06,salin06}. One
rationalization for this approximation has been that the
critical bond activation, that will irreversibly lead to
dissociation or the reflection back into the gas-phase,
often occurs at short interaction times, for which substrate
mobility is not yet a critical issue. Obviously, the actual
dynamics that proceeds after such a critical activation and
closer to the surface is then described wrongly -- the more,
the heavier the adsorbate. But this has less consequences
for the sticking probability as long as only the distinction
``direct dissociative adsorption" or ``immediate back-scattering"
matters. The situation is less clear in case of molecular
trapping that typically involves long interaction times. We
return to this problem in \sref{s:md} below.

\subsection{Adiabatic PES from DFT calculations \label{s:dft}}

\begin{figure}
\centering
\includegraphics{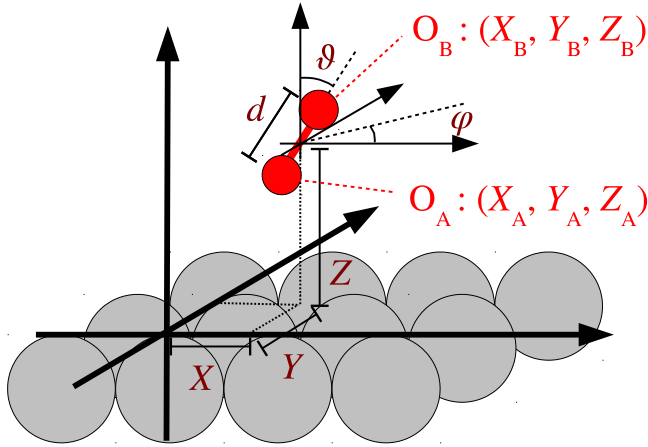}
\caption{(Colour online) Coordinate system used in the
mapping of the 6D O$_2$/Ag(111) PES. Surface Ag atoms are
shown as large grey spheres, and the O$_2$ molecule as small
red spheres. The origin of the coordinate system is placed on
a Ag surface atom.}\label{coords}
\end{figure}

Within the frozen-surface approximation the remaining
six-dimensional PES only depends on the degrees of freedom of
the O$_2$ molecule, given by the cartesian coordinates
$ \mathvec{R}_\mathid{A} = \left( X_\mathid{A},Y_\mathid{A},Z_\mathid{A} \right) $
and
$ \mathvec{R}_\mathid{B} = \left( X_\mathid{B},Y_\mathid{B},Z_\mathid{B} \right) $
of the two constituent oxygen atoms, $\mathid{A}$ and
$\mathid{B}$, respectively. In the following, another very
common equivalent coordinate system as depicted in \fref{coords}
is made use of: Molecular configurations are conveniently
described by the centre of mass of the molecule $(X,Y,Z)$, the
interatomic distance $d$, and the orientation relative to the
surface, defined by the polar and azimuth angles $\vartheta$
and $\varphi$, respectively. The origin is placed on top of a
silver atom within the surface plane, and $(\vartheta,\varphi)
= (90^{\circ}, 0^{\circ})$ corresponds to the molecular axis
oriented parallel to the $[1\bar{1}0]$ direction. To construct the
6D PES, we interpolate with the NN a large set of energies
that have been first calculated with density-functional theory
(DFT) for different positions and orientations of the molecule
over the surface. Note that the role of subsurface oxygen
\cite{todorova02,xu05} is not considered in this work because
we are interested in the measured initial sticking coefficient,
i.e. in the dissociation probabilities that are measured at
very low coverages (about 0.1~ML or less).  We have performed
more than 7000 spin-polarized DFT total energy calculations,
starting with so-called elbow plots, i.e. $(Z,d)$ cuts through
the 6D PES, at high symmetry sites of the surface. Additional
irregularly spaced molecular configurations, typically in form
of ``scans'' both in lateral $(X,Y)$ and angular $(\vartheta,
\varphi)$ dimensions, were added iteratively as needed during
the neural network interpolation (\emph{vide infra}).

An energy cut-off of 400\,eV in the plane-wave basis set
together with ultra-soft pseudopotentials as contained in the
default old Cambridge library of the CASTEP code~\cite{clark05}
have been used to describe the interaction between electrons
and nuclei, relying on the exchange-correlation functional due
to Perdew, Burke and Ernzerhof (PBE)~\cite{perdew96}. Based on
these computational settings, the computed bulk Ag lattice
constant and bulk modulus compare well to experimental
values~\cite{staroverov04} ( $a_\mathid{0,PBE}^\mathid{Ag}
= 4.14$\,{\AA} and $B_\mathid{0,PBE}^\mathid{Ag}
= 97.1$~GPa vs.  $a_\mathid{0,exp}^\mathid{Ag} =
4.07$\,{\AA} and $B_\mathid{0,exp}^\mathid{Ag} = 109$~GPa,
respectively), and also the free O$_2$ bond-length is
with $d_\mathid{0,PBE}^\mathid{O_2} = 1.24$\,{\AA} well
described ($d_\mathid{0,exp}^\mathid{O_2} = 1.21$,{\AA}
~\cite{bookami}). Notwithstanding, similar to other frozen-core 
calculations, the O$_2$ gas-phase binding energy is underestimated 
compared to accurate full-potential PBE results~\cite{kiejna06}. 
With 5.64~eV this fortuitously yields
a value close to the experimental gas-phase data~\cite{bookcrc}
(5.12~eV). For the binding energies addressed below this is
not to be of concern, as this underestimation largely cancels
in the total energy differences constituting the latter, and
a potentially introduced constant shift does not affect the
dynamics on the interpolated PES.

The Ag(111) surface is modelled with a periodic five-layer
slab separated by 12\,{\AA} of vacuum. In order to sufficiently
suppress spurious interactions of the impinging O$_2$ molecule
with its periodic images in the supercell geometry we are forced
to use a $(3 \times 3)$ surface unit cell. The irreducible
wedge of the first Brillouin zone corresponding to this
large unit cell is then sampled with a $(4\times4\times1)$
Monkhorst-Pack grid of special $k$-points. Systematic
convergence tests focusing on the adsorption of atomic and
molecular oxygen at high-symmetry sites confirm that the
binding energies are numerically converged to within 50\,meV
at these settings. For the clean Ag(111) surface we reproduce
the established marginal 1\% outward relaxation of the first
interlayer distance \cite{li02}. Within the frozen-surface
approximation we neglect this and compute the O$_2$/Ag(111) PES
above a bulk-truncated slab. As discussed in detail by Behler
{\em et al.} for O$_2$ on Al(111) \cite{behler05,behler08},
also in this adiabatic PES the spin of the O$_2$ molecule is
gradually quenched from the triplet ground-state of O$_2$
at large $Z$ distances to the spin unpolarized state close
to the surface. We use the configuration where the O$_2$
molecule is located in the middle of the vacuum region as zero
reference. This corresponds to the full decoupling limit between
O$_2$ and Ag(111), since at these distances, unlike at Al(111)
\cite{behler05,behler08}, there is no spurious charge transfer
to the molecule. In the employed sign convention negative
PES values indicate exothermicity with respect to this limit,
while positive values indicate endothermicity.

\subsection {Interpolation by symmetry--adapted neural
networks\label{s:nn}}

Neural networks represent a very flexible class of
composed functions that in principle allows to approximate
(physical) potential energy surfaces to arbitrary accuracy
\cite{cybenko89,hornik89}. However, no physical properties
of the latter are built therein \emph{a priori}. The use
of NNs in the context of gas-surface dynamics, pioneered by
Lorenz \emph{et al.} \cite{lorenz04}, has been described in
detail before \cite{lorenz06,behler07}. Therefore, we only
give a brief account of the novel aspects of how symmetry
is taken into account for the present system. Instead
of using the ``straightforward physical'' coordinates
depicted in \fref{coords}, the neural network interpolation
is performed on a properly transformed set, which has been
termed symmetry functions before~\cite{behler07}. This way,
two molecular configurations that are equivalent by symmetry
are enforced to have same energy on the NN-interpolated PES by
construction. By the rigid incorporation of these important
physical properties of a specific system, the interpolation
problem is thus simplified considerably (i.e. better quality
with less input data). The construction of such a coordinate
transformation is a rather complicated task, as it has to
capture both periodic and point group symmetry on high-symmetry
sites of the surface correctly. This implies certain demands
for the behaviour under a certain (discrete but infinite)
set of transformations including the molecular centre of mass
position on the surface $(X,Y)$ and the molecular orientation
$(\vartheta,\varphi)$, but not the distance from the surface
$Z$ and the internuclear distance $d$. Artificial symmetries
have been introduced unintentionally in past attempts and
only recently been found to have a significant effect on the
calculated sticking probability~\cite{gross11}.

\begin{figure}
\centering
\includegraphics{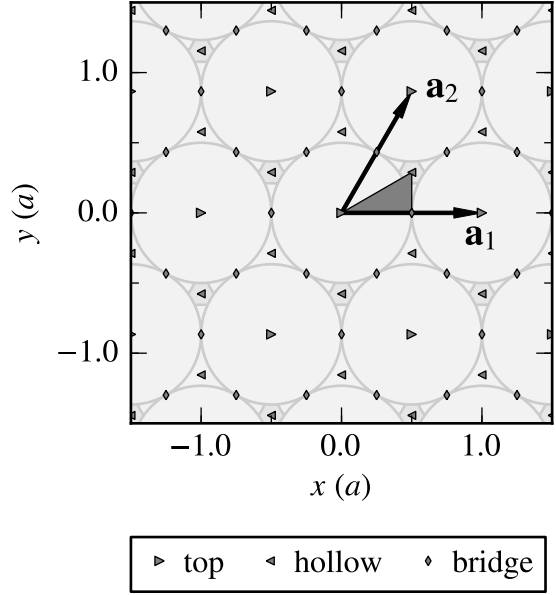}
\caption{(Colour online) Structure and sixfold symmetry of the
"(111)" surface employed in the symmetry-adapted NN approach
(see text). The primitive lattice vectors $\mathvec{a}_{1}$
and $\mathvec{a}_{2}$ are shown as arrows, high symmetry
sites are marked by different symbols, and the irreducible
wedge is indicated by the dark gray triangle. Distances are
conveniently described in units of the surface lattice constant
$a$.}\label{fcc111sixfold}
\end{figure}

For the present system, symmetries resulting from the
combination of the threefold symmetry of the (111) surface of
the closed-packed fcc metal silver and the exchange symmetry
of the homonuclear O$_2$ molecule can be further increased
by exploiting the near degeneracy of the O$_2$ interaction
with the fcc and hcp hollow sites: For equivalent molecular
configurations at these two sites, we observe a statistical
relation in our DFT data set with a root mean square error
(RMSE) of less than 5\,meV and a mean absolute deviation of
less than 3\,meV. Similar results have been obtained before for
H$_{2}$ on Pt(111)~\cite{olsen02}. For the present work we will
neglect this small difference and thus treat the fcc and hcp
sites as equivalent. This increases the symmetry of the surface
from three- to sixfold, as illustrated by the high-symmetry
points and irreducible wedge in \fref{fcc111sixfold}, together
with the primitive lattice vectors $\mathbf{a}_{1}$ and
$\mathbf{a}_{2}$. Inspired by the symmetry functions obtained
for a fcc(111) surface with threefold symmetry~\cite{behler07},
the following set of symmetry-adapted coordinates Q$_{i}$
have been designed and carefully verified not to yield
any artificial symmetries as ``side effects''. A detailed
derivation and their simple generalization to other common
low-index surfaces of both close-packed fcc and hcp metals
will be detailed elsewhere \cite{meyer12}:
\numparts
\begin{eqnarray}
\fl Q_1 & \; = \;
    \frac{1}{2} \,
    \left[ \exp\left(-\frac{Z_\mathid{A}}{2}\right) \cdot g_{1}(X_\mathid{A},Y_\mathid{A})
      \; + \;
           \exp\left(-\frac{Z_\mathid{B}}{2}\right) \cdot g_{1}(X_\mathid{B},Y_\mathid{B}) \right] \\
\fl Q_2 & \; = \;
    \exp\left(-\frac{Z_\mathid{A}}{2}\right) \cdot g_{1}(X_\mathid{A},Y_\mathid{A})
    \; \cdot \;
    \exp\left(-\frac{Z_\mathid{B}}{2}\right) \cdot g_{1}(X_\mathid{B},Y_\mathid{B}) \\
\fl Q_3 & \; = \;
    \frac{1}{2} \,
    \left[ \exp\left(-\frac{Z_\mathid{A}}{2}\right) \cdot g_{2}(X_\mathid{A},Y_\mathid{A})
           \; + \;
           \exp\left(-\frac{Z_\mathid{B}}{2}\right) \cdot g_{2}(X_\mathid{B},Y_\mathid{B}) \right] \\
\fl Q_4 & \; = \;
    \exp\left(-\frac{Z_\mathid{A}}{2}\right) \cdot g_{2}(X_\mathid{A},Y_\mathid{A})
    \; \cdot \;
    \exp\left(-\frac{Z_\mathid{B}}{2}\right) \cdot g_{2}(X_\mathid{B},Y_\mathid{B}) \\
\fl Q_5 & \; = \;
    \exp\left(-\frac{Z}{2}\right) \cdot g_\mathid{1}(X,Y) \\
\fl Q_6 & \; = \;
    \exp\left(-\frac{Z}{2}\right) \cdot g_\mathid{2}(X,Y) \\
\fl Q_7 & \; = \; \exp\left(-\frac{Z}{2}\right) \\
\fl Q_8 & \; = \; d \\
\fl Q_9 & \; = \; \left[ \cos(\vartheta) \right]^2
\quad
\end{eqnarray}
\endnumparts
$g_{1}$ and $g_{2}$ are functions $\mathbb{R}^{2} \rightarrow
\mathbb{R}$ with the periodicity given by the surface lattice
vectors $a_1$ and $a_2$.  They have been constructed such
that points $\mathvec{r} = (x,y)$ in the surface plane
which are equivalent by the sixfold symmetry depicted in
\fref{fcc111sixfold} are mapped onto the same pair of function
values
$\left(g_{1}(x,y),\, g_{2}(x,y)\right)$ \cite{meyer12}:
\numparts
\begin{eqnarray}
\fl\eqalign{
  g_{1}(x,y) & = \frac{1}{4}
    \bigg[
      \cos \bigg( \underbrace{ \frac{2\pi}{a} x - \frac{2\pi}{\sqrt{3}a} y }_{} \bigg) +
      \cos \bigg( \underbrace{ \frac{4\pi}{\sqrt{3}a} y }_{} \bigg) +
      \cos \bigg( \underbrace{ \frac{2\pi}{a} x + \frac{2\pi}{\sqrt{3}a} y}_{} \bigg) +
      1
    \bigg] \\
  & \quad\qquad\qquad \mathvec{G}_{10} \mathvec{\cdot} \mathvec{r}
    \quad\qquad\qquad \mathvec{G}_{01} \mathvec{\cdot} \mathvec{r}
    \quad\qquad\qquad \mathvec{G}_{11} \mathvec{\cdot} \mathvec{r}
} \\[1.5em]
\fl\eqalign{
  g_{2}(x,y) & = \frac{1}{4}
   \bigg[
      \cos \bigg( \underbrace{ \frac{2\pi}{a} x - \frac{2\sqrt{3}\pi}{a} y }_{} \bigg) +
      \cos \bigg( \underbrace{ \frac{4\pi}{a} x }_{} \bigg) +
      \cos \bigg( \underbrace{ \frac{2\pi}{a} x + \frac{2\sqrt{3}\pi}{a} y }_{} \bigg) +
      1
    \bigg] \\
  & \quad\qquad\qquad \mathvec{G}_{1\bar{1}} \mathvec{\cdot} \mathvec{r}
    \quad\qquad\qquad \mathvec{G}_{21} \mathvec{\cdot} \mathvec{r}
    \quad\qquad\qquad \mathvec{G}_{12} \mathvec{\cdot} \mathvec{r}
    \quad
}
\end{eqnarray}
\endnumparts
$a$ is the surface lattice constant given by $1/\sqrt{2} \,
a_\mathid{0,PBE}^\mathid{Ag}$ for the present system. The
building blocks of $g_{1}$ and $g_{2}$ are low order terms of
the corresponding Fourier series as indicated by the respective
reciprocal lattice vectors
\begin{equation}
  \mathvec{G}_{nm} = n \mathvec{b}_{1} + m \mathvec{b}_{2} \; .
\end{equation}
Here $\mathvec{b}_{1}$ and $\mathvec{b}_{2}$ are the primitive vectors
of the reciprocal lattice given by the defining relation
\begin{equation}
  \mathvec{a}_{i} \cdot \mathvec{b}_j = 2\pi \delta_{ij}
  \quad i,j \in \{1,2\} \; ,
\end{equation}
and the crystallographic convention to indicate a negative
direction by a bar over the corresponding number is adopted.

\begin{figure}
\centering
\includegraphics{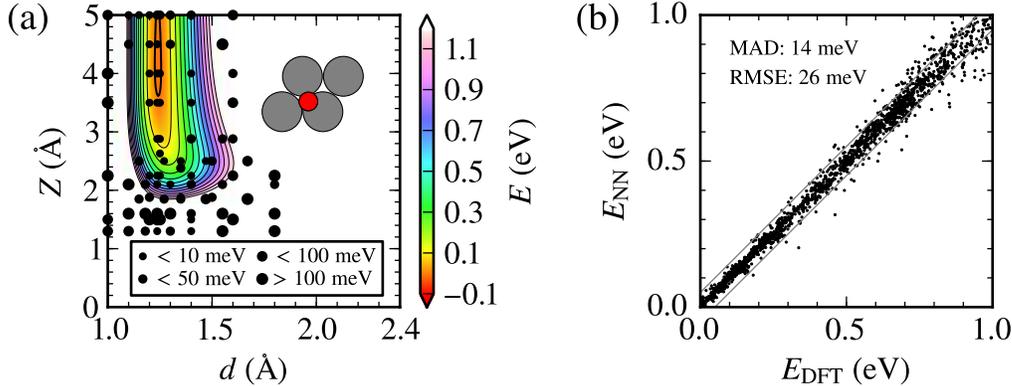}
\caption{(Colour online) Illustration of the interpolation quality obtained
with the best NN fit for the present system. (a) Representative $(d,Z)$ (elbow)
plot corresponding to a molecule with $X=0.5 \,a$, $Y=\sqrt{3}/6 \,a$ and
$\vartheta=0^{\circ}$, where $a$ is the surface lattice constant. The size of
the black circles shows the maximum error of the NN interpolation compared to
the DFT input at the corresponding configurations. (b) Comparison of the NN
interpolated values with all available corresponding DFT energies. Chosen to
match the estimated accuracy of the latter (see text), the grey lines at $\pm
50$\,meV serve as a guide to the eye for the interpolation error. The resulting
maximum absolute deviation (MAD) and root mean square error (RMSE) are given.}

\label{DFT-NN}
\end{figure}

Along the lines of previous work \cite{behler07}, neural network
training has been simplified by limiting highly repulsive
energies to $\le 5\,\mathid{eV}$, which is still much larger
than the energy range of interest for the ensuing molecular
dynamics trajectories discussed below.  Using the adaptive
extended Kalman filter algorithm including the modifications
described in \cite{lorenz06} for training, the sensitivity
of the NN is further focused on the most relevant parts of
the PES by assigning training weights
\begin{equation} \fl
\omega_\mathid{train} = \cases{
    5000 \cdot \exp \left[ -3 \left(E_\mathid{DFT}/\mathid{eV}
    + 0.1\right) \right] & for $E_\mathid{DFT} < 2\,\mathid{eV}$
    \\ 9 & for $E_\mathid{DFT} \ge 2\,\mathid{eV}$ \\}
\end{equation}
to the input data. More than forty
different NN topologies have been tried. In each case,
the interpolation quality has been carefully monitored
by plotting numerous two-dimensional $(d,Z)$, $(X,Y)$ and
$(\vartheta,\varphi)$ cuts through the such obtained PESs,
including information about the interpolation errors of
individual data points. An example for a corresponding elbow
is shown in \fref{DFT-NN}~(a). With increasing knowledge
gained about the PES and dynamics thereon, we decided for a
\{9-16-16-16-1 $tttl$\} topology of the NN. In this commonly
used notation \cite{behler08,lorenz04,lorenz06,behler07}, the
first number refers to the nine symmetry-adapted coordinates
$(Q_1,\ldots,Q_9)$ in the input layer. The three numbers in
the middle denote the number of nodes in the hidden layers,
and the last number indicates that there is a single node
yielding the PES value in the output layer. Like in previous
work \cite{behler08,lorenz06}, a hyperbolic tangent and
a linear function serve as activation functions in the
hidden and output layer and are denoted by $t$ and $l$,
respectively.
\Fref{DFT-NN}~(b) summarizes the high quality of the finally
obtained interpolation. For statistically significant subsets
of the data points in the dynamically relevant entrance channel
and chemisorption well ({\em vide infra}), the RMSE is smaller
than 26\,meV. Considering points from the training and test sets
separately, RMSEs for the latter are larger by only a few meV --
thus indicating good predictive properties for arbitrary molecular
configurations to be encountered during the dynamics.

\subsection{Molecular dynamics simulations\label{s:md}}

On the thus obtained continuous NN-representation of the
adiabatic PES we perform trajectory calculations using a
Bulirsh-Stoer integrator \cite{bulirsch66} for the classical
equations of motion in Cartesian coordinates.  Trajectories
start with the O$_2$ molecule at its calculated gas-phase bond
length and at $Z=7$\,{\AA} above the surface, neglecting its
initial zero point energy. A conventional Monte Carlo procedure
is used to sample all possible initial O$_2$ orientations
$(\vartheta, \varphi)$ and lateral positions $(X,Y)$. For each
incident energy $E_i$, the sticking coefficient is obtained
based on $10^7$ different trajectories at perpendicular
incidence.\footnote{Results of calculations for various
non-perpendicular incidence angles focusing on scattering
will be reported in a forthcoming publication.}
Such a large number of trajectories can only be integrated
because the cost to compute forces from the NN PES-representation
is negligible
compared to the DFT calculations.  As outcome of each individual
trajectory we consider the following possibilities:
\begin{enumerate}
\item The trajectory is assumed to yield a dissociation event
when the O$_2$ bond length $d$ is stretched beyond 2.4\,{\AA},
i.e. twice the gas-phase distance, and the associated velocity
$\dot{d}$ is still positive. This is a rather conservative
criterion and we have verified that some variation of the
critical bond distance has no effect on the obtained sticking
curves.
\item The molecule is assumed to be reflected when its centre
of mass reaches the initial starting distance of 7\,{\AA}
with a positive $Z$-velocity. The results were equally found
to be insensitive to reasonable variations of this criterion.
\item
Molecular trapping, when the molecule is neither dissociated
nor reflected after 24~ps. A detailed analysis of the obtained
trajectories shows that this time span is well separated from
those of reflection or dissociation events, which happen in
the order of 2--3~ps, such that the precise value of 24~ps
has little influence on the number of determined trapping
events. Although one may expect that the longer the molecule
spends close to the surface, the more probable would be energy
dissipation into either electronic or phononic excitations that
in turn would prevent these trapped molecules to eventually
escape from the surface, no final statement can be done in such
cases within the frozen surface approximation ({\em vide infra}).
\end{enumerate}

\section{Results and discussion\label{s:results}}

\subsection{Insight from experiment}

With silver as the almost unique industrial
catalyst employed in the selective epoxidation of
ethylene~\cite{campbell84,bukhtiyarov03}, the interaction
of O$_{2}$ with silver surfaces has already been extensively
investigated during the last decades in an attempt to understand
the fundamentals behind such exceptional catalytic functionality
(e.g. see \cite{michaelides05} and references therein).
The richness and complexity of the O$_2$ interaction with silver
is reflected in a variety of adsorbed species that have been
identified even at just the flat low-index surfaces Ag(100),
Ag(110), and Ag(111), when varying the exposure conditions such
as surface temperature, oxygen coverage, gas temperature and gas
pressure~\cite{campbell84,campbell85,grant84,schmeisser85,bartolucci}. In
ultra-high vacuum physisorption states are observed at lowest
temperatures, while molecularly chemisorbed states prevail
up to $T_s \sim 150$~K. Dissociative adsorption appears for
higher crystal temperatures; in this case subsurface oxygen
can be also important for coverages where silver oxides are
about to be formed \cite{todorova02}. All these studies showed
that the abundance of each state also depends on the surface
structure, with the close-packed Ag(111) surface being the
most inert towards thermal oxygen adsorption~\cite{campbell85}.

All molecular beam experiments performed on the three
low-index faces show that, if present, both the molecular
and the dissociative initial adsorption probabilities are always
activated~\cite{vattuone94jcp,vattuone94prl,raukema96jpcm,vattuone98,raukema96ss}.
Despite this consensus, there remain unresolved experimental
discrepancies on the Ag(111) surface regarding the actual
order of magnitude of the dissociation probability at normal
incidence. Using a molecular beam, Raukema {\em et al.} reported
initial dissociation probabilities that increase from $10^{-5}$
to $10^{-3}$ as the incidence energy $E_i$ increases in the
range 0.8--1.7~eV~\cite{raukema96ss}. In contrast, similar
experiments performed with a molecular beam of $0.098 < E_i <
0.8$~eV, estimated that the dissociation probability on the {\em
clean} Ag(111) surface should be lower than $9\times10^{-7}$ at
least in that $E_i$-range~\cite{rocca97}. This has been related
\cite{rocca97} to the particular propensity of the Ag(111)
surface towards surface imperfections, notably steps, or towards
residual gas contamination at lower surface temperatures.

In this situation the main purpose of our modelling is
evidently not so much to reach full quantitative agreement with
experiment, but rather to establish the order of magnitude
for the initial dissociation probability within an adiabatic
picture. As we will show below this is already sufficient to
reach important conclusions concerning the role of possible
non-adiabatic effects. In addition it contributes to the
discussion of possible sources for the differences between the
experimental data sets of~\cite{raukema96ss,rocca97} and will
provide a valuable basis for future measurements that aim to
overcome the present scatter in the experimental data.

\subsection{Adiabatic PES properties\label{s:pes}}

\begin{figure}
\centering
\includegraphics{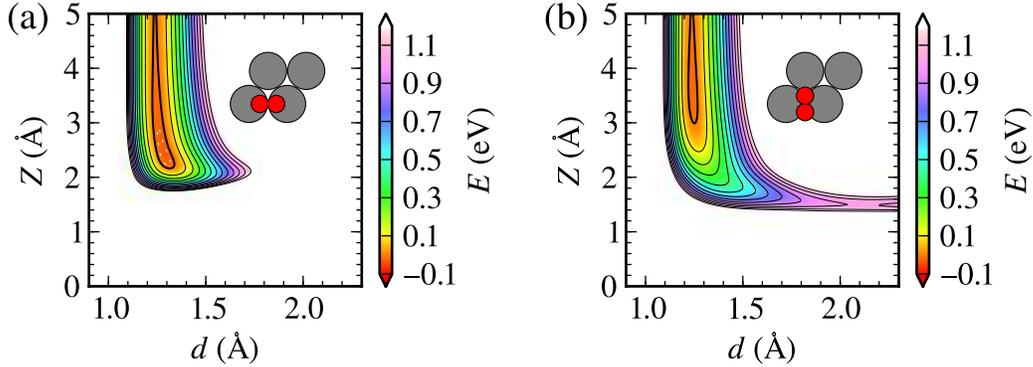}
\caption{(Colour online) Elbow plots of two relevant
configurations of the O$_2$/Ag(111) PES: (a) The chemisorption
state in which the molecule is located over bridge and with
$(\vartheta=90^{\circ}; \varphi=0^{\circ})$. (b) Configuration
for the minimum energy barrier to dissociation in which the
molecule is located over bridge with $(\vartheta=90^{\circ};
\varphi=90^{\circ})$.}\label{2dcuts}
\end{figure}

Overall, the obtained O$_2$/Ag(111) PES is rather repulsive. A
systematic analysis of $(d,Z)$ cuts for different molecular
configurations shows that at intermediate distances from the
surface ($Z \approx 2$\,{\AA}) the potential energy is already
above one eV in almost all cases. In this respect, the $(d,Z)$
cuts of \fref{DFT-NN}~(a) and \fref{2dcuts}~(a) are rather
representative. As expected, the repulsion is strongest over
the top sites. Already at distances $Z > 3$\,{\AA} far from the
surface the emerging corrugation gives rise to a very shallow
well with a depth of the order of 20-30\,meV. As the employed
semi-local PBE functional does not account for van der Waals
interactions, this is unlikely a proper representation of the
experimentally reported physisorption state, but instead rather
an artefact of the NN interpolation. As we will discuss below,
this shallow (and likely spurious) well only plays a role for
the O$_2$/Ag(111) gas-surface dynamics at lowest incidence
energies and its effects on the dissociation probability are
easily separated.

Much more relevant for the dynamics is instead a second
well that we determine at $Z=2.3$\,\AA, and for a molecular
configuration where the O$_2$ centre of mass is situated above
a bridge site and with the molecular axis pointing to the
two closest Ag atoms. \Fref{2dcuts}~(a) shows the elbow plot
corresponding to this configuration. The computed internuclear
distance is with $d = 1.28$\,{\AA} only slightly stretched
compared to the gas-phase value ($d = 1.24$\,{\AA}). This
indicates a rather weak interaction that we find equally
reflected in the computed low binding energy of $-40$\,meV
and the modest shift of the O$_2$ stretch vibration from the
computed 192\,meV gas-phase value down to 145\,meV. These
properties are only little affected by the frozen surface
approximation and the NN interpolation. After a direct
DFT calculation of this state that also allows for surface
relaxation the most notable change is the increase of the
binding energy to $-70$\,meV. The latter value is somewhat
lower than the $-0.17$\,eV binding energy determined before
by Xu {\em et al.} for the same O$_2$ molecular adsorption
state \cite{xu05}. Considering the highly similar computational
setup employed in the latter study we attribute this $0.1$\,eV
difference primarily to their use of a smaller $(2 \times 2)$
surface unit-cell.

It is tempting to identify this molecular state with
the chemisorption state reported in the experimental
literature. For this, one has to recognize though that
(regardless of the small scatter) both theoretical binding
energies are at variance with the experimental estimate
for the chemisorption state of $\sim 0.4$\,eV from thermal
desorption spectroscopy (TDS) \cite{campbell85}. This
difference is noteworthy as preceding equivalent DFT
calculations performed for O$_2$ at Ag(100)~\cite{alducin08} and
Ag(110)~\cite{gravil96,monturet10} reproduce the approximately
$-0.4$\,eV chemisorption energies measured at these facets
~\cite{campbell84,buatier97}. We also confirmed
that the here employed computational setup fully reproduces
the O$_2$ binding energy reported by Alducin {\em et al.}
at Ag(100) \cite{alducin08}. This suggests that both the
employed PW91 and PBE functionals, which are very closely
related, are capable of describing the O$_2$-Ag interaction
rather well. Similarly problematic is the observation that
none of the experimentally reported vibrational frequencies
\cite{carley90,mongeot95}
comes close to the 145\,meV that we compute for the
molecular state. Instead much lower frequencies below
100\,meV are measured, which, however, have already
been assigned to hydroxyl, water groups or imperfections
\cite{carley90,mongeot95}.
Furthermore, as shown by Xu {\em et al.} \cite{xu05} also
sub-surface oxygen can strongly stabilize chemisorbed
O$_2$ at Ag(111). In this situation we believe that the
low computed binding energy of the molecular state rather
supports the interpretation that the chemisorption state
prepared in the experiments by Raukema, Campbell and others
\cite{raukema96ss,kleyn96,campbell85,schmeisser85}
is not a property of the pristine Ag(111) surface, but reflects
the propensity of this surface towards defects or residual
gas contamination, and we will return to this point below when
discussing the dissociative sticking probability. Within the
theory-theory comparison of O$_2$ at the three low-index
surfaces, the lower binding energy computed here at
Ag(111) as compared to Ag(100) \cite{alducin08} and Ag(110)
\cite{gravil96,monturet10} is instead perfectly consistent with
the expected chemical inertness of the close-packed surface.

As a final important characteristic of the adiabatic PES we also
determined the minimum energy path (MEP) that leads out of the
molecular well to dissociation. This path goes along the $(d,Z)$
PES cut shown in \fref{2dcuts}~(b), in which the molecule is
located over a bridge site with its axis oriented along the
direction joining the closest fcc/hcp sites. The transition
state corresponds to an activation barrier of 1.1\,eV, which is
in very good agreement to the value reported before by Xu {\em
et al.} \cite{xu05}. It is located at $Z \approx 1.6$\,{\AA},
i.e. at a distance closer to the surface ($Z \sim 2-3$\,{\AA})
than where one would expect barriers due to spin or charge non-adiabaticity
({\em vide infra}).

\subsection{Dissociative sticking coefficient\label{s:s0}}

\begin{figure}
\centering
\includegraphics{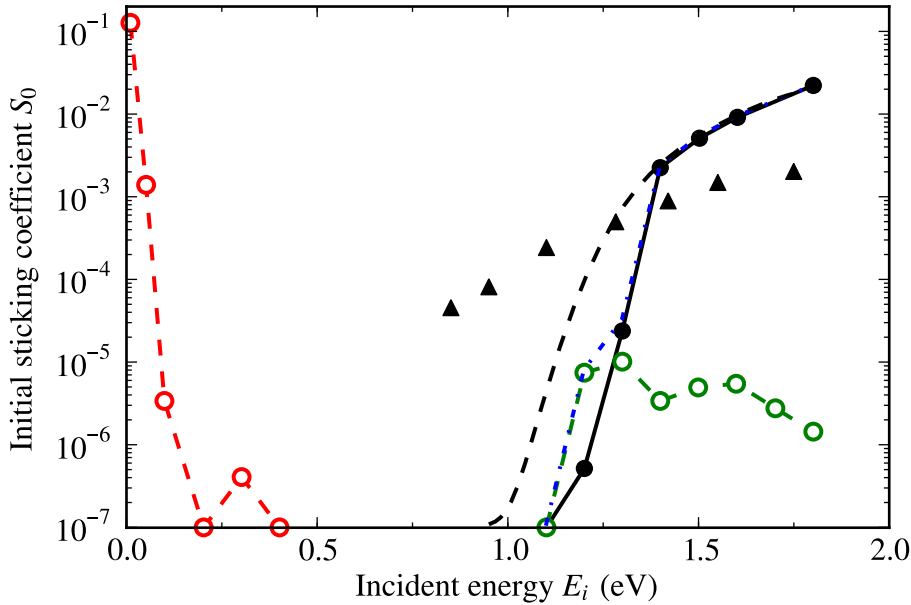}
\caption{(Colour online) Initial sticking coefficient $S_0(E_i)$
of O$_2$ on Ag(111) as a function of the incident energy $E_i$
under normal incidence conditions. Shown are the computed
probabilities for direct dissociation (black full circles),
and for molecular trapping in the low (red open circles)
and high (green open circles) energy range (see text for
details). The blue dash-dotted line is included
to illustrate the total dissociation under the assumption that
all of the latter events eventually lead to dissociation
(i.e. sum of the green open circles and the black full circles).
The black dashed line illustrates the effect of the finite
energetic width of the experimental beam on the simulated direct
dissociative sticking curve according to \eref{eq:S0corr}. The
triangle data points reproduce the experimental data of
\cite{raukema96ss} obtained at $T_s=400$~K. \label{sticking}}
\end{figure}

We compile the results of our adiabatic sticking coefficient
simulations for normal incidence in \fref{sticking}. Two
different regions can clearly be discerned. At incidence
energies below 0.4\,eV there is an appreciable amount of
molecular trapping events. This component rapidly decreases
with increasing $E_i$, as characteristic for physisorption. Note
that the non-monotonous scatter of this curve around $E_i \sim
0.2-0.4$\,eV is already no longer statistically relevant. For
the employed $10^7$ trajectories the corresponding probabilities
around $10^{-7}$ mean that we have observed either one or two
molecular trapping events. Analysis of the trajectories in this
entire low-energy trapping branch reveals that these events are
all due to dynamical trapping in the shallow PES well located
at distances $Z > 3$\,{\AA}. As we are rather sceptical about
the meaning of this shallow well we would not put too much
emphasis on  this low-energy results, even though the order of
magnitude of this trapping and the incidence energy range fit
rather well to those of the physisorption trapping discussed by 
Raukema and Kleyn \cite{raukema95,raukema95prl}. We note however
that a proper description of any kind of molecular adsorption
requires to go beyond the frozen surface approximation by
including surface temperature effects and energy dissipation
channels in the simulations. In any case, this lowest-energy
trapping part is well separated from the sticking properties
at incidence energies higher than 0.4\,eV, and for the latter
part the shallow PES well plays no further role.

This other part that we will exclusively discuss henceforth
starts at incidence energies above 1.1\,eV and is characterized
by a steeply increasing dissociative sticking probability. In
contrast, in the intermediate range 0.4\,eV$< E_i <$1.1\,eV we
have not observed a single dissociation event in the total of
$10^7$ trajectories performed for every incidence energy. From
our calculations we can thus put an upper bound to the real
dissociative sticking probability at these incidence energies
of $10^{-7}$. Above 1.1\,eV this probability increases steeply
and seems to saturate above $10^{-2}$ at the highest incidence
energies considered.  Analysis of the trajectories shows that
the dissociation follows a rather direct mechanism ruled by
the large energy barriers existing at distances below 2\,{\AA}
from the surface. Almost all of the dissociating molecules
follow paths very close to the MEP shown in \fref{2dcuts}(b)
above. This explains the coincidence of the onset of sticking
above 1.1\,eV with the activation energy along this path and
rationalizes the overall very low sticking probability at this
surface in terms of the narrow dissociation channel and thus
reduced configurational space.

If we compare these sticking results with the experimental
data from Raukema {\em et al.} \cite{raukema96ss} that is
also included in \fref{sticking} we first notice a gratifying
overall agreement in the sense that the computed adsorption
probabilities are in the right (low) order of magnitude and
above threshold show the correct monotonous increase up to the
highest incidence energies considered. However, quantitatively,
there are notable deviations. This concerns foremost a much
steeper slope of the theoretical curve especially for $E_i <
1.3$\,eV. As a consequence, there is still appreciable sticking
of the order $10^{-5} - 10^{-4}$ at 0.8-0.9\,eV incidence
energies in the experiments, whereas the theoretical curve
drops steeply to below $10^{-7}$ already at $E_i = 1.1$\,eV.

One reason that immediately comes to mind to generally rationalize such
discrepancies is the approximate nature of the employed xc functional. In the
present case, this is unlikely to explain all of the deviations though. As
described before, almost all of the direct dissociation goes along paths close
to the MEP shown in \fref{2dcuts}(b). If the xc functional messed up the MEP
barrier height, this would thus -- to zeroth order -- predominantly have
resulted in a rigidly shifted theoretical sticking curve to too high or too low
$E_i$, and less to a wrong slope. If the employed PBE functional e.g.
overestimated the true MEP barrier height, a ``better" functional would shift
the curve to lower $E_i$. This would then lead to a better agreement at the
onset region. However, simultaneously it would increase the difference to the
experimental curve at the highest incidence energies, with the theoretical
curve then seriously overshooting the experimental data there.

From this reasoning we thus rather suspect other factors to
stand behind the disagreement between the theoretical and
experimental sticking curve. The influence of vibrationally excited 
O$_2$ in the beam can be neglected, according to the low populations 
(less than $10^{-3}$ for $\nu = 3$ and than $10^{-2}$ for $\nu = 2$)
that have been estimated by Raukema et al. based on the nozzle
temperature \cite {raukema96ss}. Another factor that could be
important in particular in light of the abrupt initial increase
of the theoretical sticking curve is the finite energetic
width of the experimental beam \cite{martin-gondre11}. To mimic
the effect of the latter in the theoretical data we recompute
the curve simply as a convolution with an energy distribution
function $f_{\beta,E_{0}}(E)$ for the beam:
\begin{equation}
\label{eq:S0corr}
  S_{0}^{\rm corr}(E) =
    \int\limits_{0}^{\infty} \rmd {E'} \; \tilde{S}_0(E') \; f_{\beta,E'}(E)
\end{equation}
This equation is evaluated numerically, and $\tilde{S}_0(E')$
describes the corresponding curve plotted in \fref{sticking}
(i.e. a piecewise linear interpolation based on the explicitly
calculated points). $f_{\beta,E_{0}}(E)$ is taken to be a
shifted Maxwell-Boltzmann distribution
\begin{equation}
\label{eq:fbeam}
  f_{\beta,E_{0}}(E) =
    \left[ N(\beta,E_{0}) \right]^{-1} \;
    E \, \exp{\left[-\beta \left(\sqrt{E}-\sqrt{E_{0}}\right)^2 \right]}
    \quad .
\end{equation}
Formulated in the velocity domain, it is known to properly describe supersonic
molecular beams \cite{haberland85} and has also been used by Raukema
for the evaluation of time-of-flight (TOF) experiments
for the present system \cite{raukema96diss}.
$ N(\beta,E_{0}) = \int_{0}^{\infty} \rmd {E'} \; f_{\beta,E_{0}}(E') $
is a normalization factor, and the width parameter $\beta$ is chosen to
reproduce the experimental full width at half maximum (FWHM) of 0.2~eV in the
energy range of interest \cite{raukema95}. As seen in \fref{sticking},
this correction leads to a smoother theoretical curve that has its onset
with 0.9\,eV now at slightly lower incidence energies.
Nevertheless, it does not help to reconcile experiment and theory especially at
these $E_i$ close to threshold, where the experiment by Raukema {\em et al.}
\cite{raukema96ss} yields a dissociative sticking that is more than two orders
of magnitude higher than the computed one. Our results are instead more
compatible to those of \cite{rocca97}, which estimated a $S_0(E_i)$ below
$9\times10^{-7}$ for $E_i = 0.8$\,eV.

Another factor that should be kept in mind is the finite surface
temperature $T_s$ in the experiments that is not accounted
for at all in the present theoretical approach. For the normal
incidence data shown in \fref{sticking} Raukema {\em et al.}
used $T_s = 400$\,K \cite{raukema96ss}. For larger incidence
angles they also reported data measured at a lower $T_s =
220$\,K. There, the lowering of the surface temperature led
to an increase of the slope of the sticking curve above $E_i
= 1.0$\,eV \cite{raukema96ss}. Unfortunately, no such lower
temperature data was presented for normal incidence. If we
assume that the trend seen at larger incidence angles prevails,
we would expect normal incidence measurements at lower $T_s$
to yield a larger slope than the one of the experimental $T_s =
400$\,K curve shown in \fref{sticking} -- in closer agreement
to the calculated sticking data.

In this respect it is also important to note that from the
observed non-monotonous dependence of $S_0(E_i)$ on $T_s$ and on
the incidence conditions ($E_i$, incidence angle) Raukema {\em
et al.} proposed the existence of two distinct dissociation
mechanisms~\cite{raukema96ss}:
\begin{enumerate}
\item a direct one that governs
dissociation at normal incident energies $E_{i} >1$\,eV and
\item a precursor-mediated one at work in the threshold energy
regime ($E_{i} <1$~eV), in which a molecular chemisorption state
acts as the precursor.
\end{enumerate}
In the absence of substrate mobility
and concomitant phononic dissipation channels in the model,
the closest we can get to a precursor-mediated mechanism within
the present theoretical approach are the molecularly trapped
events we indeed observe for $E_i >1.1$\,eV. The probability for
such trapping is rather energy independent and about $10^{-6}
- 10^{-5}$. Even if we thus assume each of these trajectories
to eventually lead to dissociation, at this magnitude this
contribution only significantly affects the total sticking
coefficient at the direct onset around $E_i = 1.1$\,eV as
shown in \fref{sticking}.

The resulting kinked sticking curve then has a shape that
is very reminiscent of the experimental curve measured by
Raukema {\em et al.} \cite{raukema96ss} for an off-normal
incidence angle of $30^\circ$ and low surface temperature
$T_s = 220$\,K. Unfortunately, again for normal incidence
the reported data does not extend to similarly low incidence
energies and surface temperatures to resolve this kink. Vice
versa, we can at present only speculate if an account of
substrate mobility in the calculations would for $T_s =
400$\,K modify the present {\em frozen surface} total sticking
curve in \fref{sticking} (dissociation plus trapping)
to such an extent to reach quantitative agreement with
the shown experimental curve. Particularly because of our
scepticism that the here computed molecularly bound state on
the adiabatic PES corresponds to the experimentally reported
molecular chemisorption state, we do not believe that this
will be the case. From the dependence on surface temperature
Raukema {\em et al.} estimate a trapping probability of
their precursor molecular chemisorption state of the order
of $10^{-4}$ at incidence energies as low as $E_i = 0.5$\,eV
\cite{raukema96ss}. At such low energies our computed molecular
state can not be accessed at all.

Our current interpretation is therefore instead that the
discrepancy in the initial sticking coefficient below
$E_i = 1$\,eV is primarily caused by another molecular
precursor state that is absent in the theoretical
calculations and that is not a property of pristine
Ag(111). This state would then also correspond to
the strongly bound state seen in the TDS experiments
\cite{campbell84,campbell85,grant84,schmeisser85}. This
interpretation that a molecular precursor state associated to
surface imperfections is predominantly responsible for the
dissociation probabilities in the intermediate energy range
below $E_i = 1.0$\,eV has already been advocated by Rocca
{\em et al.} \cite{rocca97} to rationalize the discrepancy
in their measured sticking data and the one of Raukema {\em
et al.} precisely in this energy range. Our analysis fully
supports this view and places an upper limit of $10^{-7}$
to the dissociative sticking at ideal Ag(111) for incidence
energies up to at least $E_i = 0.9$\,eV. In this situation,
new experiments are self-evidently required to finally settle
this cause. Particularly for normal incidence they should
be performed for a wide range of incidence energies (as in
Raukema's work) and for as low surface temperatures as possible
(as to not getting riddled by residual gas contamination).

Finally, we come back to the original motivation
and the possibility that the discrepancy between
experiment and theory could also be due to non-adiabatic
effects. However, such effects that either involve sudden charge
transfer~\cite{zhdanov97} or spin flip hindrance~\cite{behler05}
tend to release the system in an excited state and would
thus in general further lower the dissociative sticking,
which is already too low in its present adiabatic form. Due
to the specific characteristics of the adiabatic PES not
even such a lowering seems likely though. In contrast to
e.g. the O$_2$/Al(111) system \cite{behler05,behler08},
the adiabatic dissociative sticking at Ag(111) is governed
by very late energy barriers larger than 1.1\,eV and at
$Z$-distances very close to the surface. Potential spin-
or charge transfer-induced hindrances instead occur earlier
in the entrance channel at $Z$-distances around 2-3\,{\AA}
\cite{behler05,behler08,zhdanov97}. Even if such hindrances
would give rise to additional barriers, those molecules that are
able to surmount them would thus still be confronted with the
adiabatic barriers when they approach closer to the surface and
into the PES region where the spin quenching or charge transfer
has then already taken place. A noticeable spin-selection rule
induced lowering of the dissociative sticking as compared to
our present adiabatic curve would therefore only occur, if the
additional restrictions in the entrance channel were larger
than those due to the late adiabatic barriers. This is not
possible for energetic reasons, because the present adiabatic
MEP barrier is with 1.1\,eV already higher than the gas-phase
O$_2$ singlet-triplet gap of 0.98\,eV \cite{bookami}. Similar
arguments hold for possible non-adiabaticity effects due to
charge transfer limitations, considering that the electron
affinity of O$_2$ is 0.44\,eV \cite{bookami}.

\begin{figure}
\centering
\includegraphics{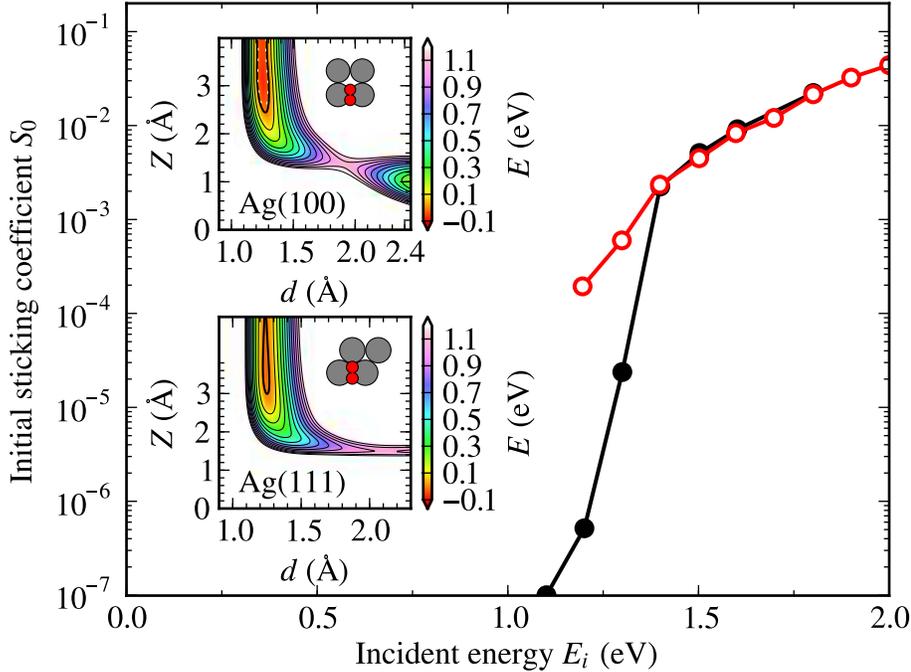}
\caption{(Colour online) Comparison of the computed adiabatic
initial dissociative sticking coefficient of O$_2$ on Ag(111)
(black full circles, copied from \fref{sticking}) and on Ag(100)
(red open circles, taken from \cite{alducin08}) for normal
incidence conditions. The insets show 2D cuts of the minimum
energy barrier configurations leading to dissociation on Ag(100)
(upper inset) and on Ag(111) (lower inset). \label{s0vsfaces}}
\end{figure}

As such we reach the same conclusions with respect
to non-adiabatic effects as for O$_2$ dissociation at
Ag(100) \cite{alducin08}. Also there, the dissociative sticking
was found to be governed by large and late barriers above 1\,eV
on the adiabatic PES. Correspondingly, the computed sticking
coefficients at Ag(100) and at Ag(111) are strikingly similar
as shown in \fref{s0vsfaces}. The higher values at Ag(100)
for lower incidence energies are due to the higher accessible
configurational space as illustrated in \fref{s0vsfaces}
by the elbow plots along the MEP. Intriguingly, at Ag(100)
experiments do confirm the absence of dissociative sticking at
0.9\,eV \cite{vattuone98}, which (considering the analogy of the
theoretical findings at the two Ag surfaces) indirectly gives
further support to our assessment that surface imperfections
cause the larger dissociative sticking around of $10^{-4}$
at this energy in the experiments by Raukema {\em et al.}
on Ag(111).

\section{Summary and conclusions}

In summary, we have employed a DFT-based divide and conquer
approach to study the reactivity of O$_2$ molecules impinging
onto the Ag(111) surface. The determined adiabatic dissociative
adsorption probability is extremely low and only exceeds
$10^{-7}$ for incidence energies above $1.1$\,eV. Our analysis
directly relates this low reactivity to large energy barriers
in excess of 1.1~eV located very close to the surface. This
excludes that electronically non-adiabatic effects, either
due to spin-selection rules \cite{behler05,behler08} or due to
charge transfer limitations \cite{citri96,zhdanov97}, contribute
to the low sticking coefficient. The late adiabatic barriers
exceed both the gas-phase O$_2$ singlet-triplet gap and the
O$_2$ electron affinity. Even if non-adiabatic effects would
give rise to additional barriers in the entrance channel, those
molecules that are able to surmount them would thus still be
confronted with higher barriers when they approach the adiabatic
PES region closer to the surface. As such, the sticking
coefficient is insensitive to a possible non-adiabaticity in
the O$_2$-Ag(111) interaction and other quantities will have
to be found as useful signatures for such effects.

The highly activated nature of O$_2$ dissociation at Ag(111) is
in good agreement with the existing experimental data. Still,
quantitatively, large deviations of more than two orders of
magnitude are obtained at intermediate incidence energies around
0.8-0.9\,eV, when taking the particular data set from Raukema
{\em et al.} \cite{raukema96ss} as reference. Instead, our data
is well consistent with the estimate of $9 \times 10^{-7}$
for this energy range by Rocca {\em et al.} \cite{rocca97},
who had already suggested surface imperfections as reason for
the higher dissociation probability in Raukema's data. The
analysis of possible uncertainties in the present approach
fully supports this point of view. At best, the high surface
temperature employed in the experiments by Raukema {\em et al.}
could be an alternative reason, and in this respect it would be
intriguing to repeat the present adiabatic calculations with a
model that includes some account of substrate mobility. However,
more important to finally settle the cause would be new
experiments at lower surface temperatures and specifically
investigating the role of surface imperfections (like the step
density) for the dissociative sticking of O$_2$ at Ag(111).

\ack We thank Prof. A. Kleyn for useful and clarifying discussions regarding
their molecular beam experiments. The work of I.G. has been supported by the
Spanish Research Council (CFM-CSIC, grant84 Nos. JAE-Pre\_08\_0045 and
2010ESTCSIC-02167). Funding by the Deutsche Forschungsgeinschaft (RE 1509/7-1)
is gratefully acknowledged. M.A. and I.J. also acknowledge the Spanish
Ministerio de Ciencia e Innovaci\'on (grant84 No. FIS2010-19609-C02-02).

\section*{References}

%\bibliographystyle{iopart-num}
%\bibliography{O2Ag111}

\begin{thebibliography}{10}
\expandafter\ifx\csname url\endcsname\relax
  \def\url#1{{\tt #1}}\fi
\expandafter\ifx\csname urlprefix\endcsname\relax\def\urlprefix{URL }\fi
\providecommand{\eprint}[2][]{\url{#2}}

\bibitem{behler05}
Behler J, Delley B, Lorenz S, Reuter K and Scheffler M 2005 {\em Phys. Rev.
  Lett.\/} {\bf 94} 036104

\bibitem{behler08}
Behler J, Reuter K and Scheffler M 2008 {\em Phys. Rev. B\/} {\bf 77} 115421

\bibitem{carbogno08}
Carbogno C, Behler J, Gro{\ss} A and Reuter K 2008 {\em Phys. Rev. Lett.\/}
  {\bf 101} 096104

\bibitem{carbogno10}
Carbogno C, Behler J, Reuter K and Gro{\ss} A 2010 {\em Phys. Rev. B\/} {\bf
  81} 035410

\bibitem{meyer11}
Meyer J and Reuter K 2011 {\em New J. Phys.\/} {\bf 13} 085010

\bibitem{vattuone94jcp}
Vattuone L, Rocca M, Boragno C and Valbusa U 1994 {\em J. Chem. Phys.\/} {\bf
  101} 713--25

\bibitem{vattuone94prl}
Vattuone L, Boragno C, Pupo M, Restelli P and Rocca M 1994 {\em Phys. Rev.
  Lett.\/} {\bf 72} 510--3

\bibitem{raukema96jpcm}
Raukema A, Butler D~A and Kleyn A~W 1996 {\em J. Phys. Cond. Mat.\/} {\bf 8}
  2247

\bibitem{vattuone98}
Vattuone L, Burghaus U, Valbusa U and Rocca M 1998 {\em Surf. Sci.\/} {\bf 408}
  L693 -- L697

\bibitem{raukema96ss}
Raukema A, Butler D~A, Box F~M~A and Kleyn A~W 1996 {\em Surf. Sci.\/} {\bf
  347} 151--68

\bibitem{chesters75}
Chesters M~A and Somorjai G~A 1975 {\em Surf. Sci.\/} {\bf 52} 21--8

\bibitem{canning84}
Canning N~D~S, Outka D and Madix R~J 1984 {\em Surf. Sci.\/} {\bf 141} 240--54

\bibitem{deng05}
Deng X, Min B~K, Guloy A and Friend C~M 2005 {\em J. Am. Chem. Soc.\/} {\bf
  127} 9267--70

\bibitem{rocca97}
Rocca M, Cemic F, de~Mongeot F~B, Valbusa U, Lacombe S and Jacobi K 1997 {\em
  Surf. Sci.\/} {\bf 373} 125--6

\bibitem{citri96}
Citri O, Baer R and Kosloff R 1996 {\em Surf. Sci.\/} {\bf 351} 24--42

\bibitem{zhdanov97}
Zhdanov V~P 1997 {\em Phys. Rev. B\/} {\bf 55} 6770--2

\bibitem{alducin08}
Alducin M, Busnengo H~F and Mui{\~n}o R~D 2008 {\em J. Chem. Phys.\/} {\bf 129}
  224702

\bibitem{bookami}
Radzig A~A and Smirnov B~M (eds) 1985 {\em Reference Data on Atoms, Molecules
  and Ions\/} (Berlin: Springer)

\bibitem{kleyn96}
Kleyn A~W, Butler D~A and Raukema A 1996 {\em Surf. Sci.\/} {\bf 363} 29--41

\bibitem{gross10}
Gro{\ss} A 2010 {\em ChemPhysChem\/} {\bf 11} 1374--81

\bibitem{engdahl92}
Engdahl C, Lundqvist B~I, Nielsen U and N{\o}rskov J~K 1992 {\em Phys. Rev.
  B\/} {\bf 45} 11362--5

\bibitem{gross95}
Gro{\ss} A, Wilke S and Scheffler M 1995 {\em Phys. Rev. Lett.\/} {\bf 75}
  2718--21

\bibitem{wiesenekker96}
Wiesenekker G, Kroes G~J and Baerends E~J 1996 {\em J. Chem. Phys.\/} {\bf 104}
  7344--58

\bibitem{diaz09}
D{\'i}az C, Pijper E, Olsen R~A, Busnengo H~F, Auerbach D and Kroes G~J 2009
  {\em Science\/} {\bf 326} 832

\bibitem{geethalakshmi11}
Geethalakshmi K~R, Juaristi J~I, Mui{\~n}o R~D and Alducin M 2011 {\em Phys.
  Chem. Chem. Phys.\/} {\bf 13} 4357--64

\bibitem{busnengo00}
Busnengo H~F, Dong W and Salin A 2000 {\em Chem. Phys. Lett.\/} {\bf 320}
  328--34

\bibitem{ischtwan94}
Ischtwan J and Collins M~A 1994 {\em J. Chem. Phys.\/} {\bf 100} 8080--8

\bibitem{lorenz04}
Lorenz S, Gro{\ss} A and Scheffler M 2004 {\em Chem. Phys. Lett.\/} {\bf 395}
  210--5

\bibitem{lorenz06}
Lorenz S, Scheffler M and Gro{\ss} A 2006 {\em Phys. Rev. B\/} {\bf 73} 115431

\bibitem{behler07}
Behler J, Lorenz S and Reuter K 2007 {\em J. Chem. Phys.\/} {\bf 127} 014705

\bibitem{pijper01}
Pijper E, Somers M~F, Kroes G~J, Olsen R~A, Baerends E~J, Busnengo H~F, Salin A
  and Lemoine D 2001 {\em Chem. Phys. Lett.\/} {\bf 347} 277

\bibitem{alducin06}
Alducin M, Mui{\~n}o R~D, Busnengo H~F and Salin A 2006 {\em Phys. Rev.
  Lett.\/} {\bf 97} 056102

\bibitem{salin06}
Salin A 2006 {\em J. Chem. Phys.\/} {\bf 124} 104704

\bibitem{todorova02}
Todorova M, Li W~X, Ganduglia-Pirovano M~V, Stampfl C, Reuter K and Scheffler M
  2002 {\em Phys. Rev. Lett.\/} {\bf 89} 096103

\bibitem{xu05}
Xu Y, Greeley J and Mavrikakis M 2005 {\em J. Am. Chem. Soc.\/} {\bf 127}
  12823--7

\bibitem{clark05}
Clark S~J, Segall M~D, Pickard C~J, Hasnip P~J, Probert M~I~J, Refson K and
  Payne M~C 2005 {\em Z. Kristallogr.\/} {\bf 220} 567--70

\bibitem{perdew96}
Perdew J~P, Burke K and Ernzerhof M 1996 {\em Phys. Rev. Lett.\/} {\bf 77}
  3865--8

\bibitem{staroverov04}
Staroverov V~N, Scuseria G~E, Tao J and Perdew J~P 2004 {\em Phys. Rev. B\/}
  {\bf 69} 075102

\bibitem{kiejna06}
Kiejna A, Kresse G, Rogal J, Sarkar A~D, Reuter K and Scheffler M 2006 {\em
  Phys. Rev. B\/} {\bf 73} 035404

\bibitem{bookcrc}
Weast R~C, Astle J~M and Beyer W~H (eds) 1986 {\em {CRC Handbook of Chemistry
  and Physics}\/} (Florida: CRC Press)

\bibitem{li02}
Li W~X, Stampfl C and Scheffler M 2002 {\em Phys. Rev. B\/} {\bf 65} 075407

\bibitem{cybenko89}
Cybenko G 1989 {\em Math. Control, Signals, Syst.\/} {\bf 2} 303--14

\bibitem{hornik89}
Hornik K, Stinchcombe M and White H 1989 {\em Neural Networks\/} {\bf 2}
  359--66

\bibitem{gross11}
Gro{\ss} A 2011 {\em J. Chem. Phys.\/} {\bf 135} 174707

\bibitem{olsen02}
Olsen R~A, Busnengo H~F, Salin A, Somers M~F, Kroes G~J and Baerends E~J 2002
  {\em J. Chem. Phys.\/} {\bf 116} 3841--55

\bibitem{meyer12}
Meyer J and Reuter K 2012 {Symmetry Adapted Coordinates for Diatomics on
  Low-Index Surfaces} to be published

\bibitem{bulirsch66}
Bulirsch R and Stoer J 1966 {\em Numerische Mathematik\/} {\bf 8} 1--13
  10.1007/Bf02165234

\bibitem{campbell84}
Campbell C~T and Paffett M~T 1984 {\em Surf. Sci.\/} {\bf 139} 396--416

\bibitem{bukhtiyarov03}
Bukhtiyarov V~I, H\"avecker M, Kaichev V~V, Knop-Gericke A, Mayer R~W and
  Schl\"ogl R 2003 {\em Phys. Rev. B\/} {\bf 67} 235422

\bibitem{michaelides05}
Michaelides A, Reuter K and Scheffler M 2005 {\em J. Vac. Sci. Technol. A\/}
  {\bf 23} 1487--97

\bibitem{campbell85}
Campbell C~T 1985 {\em Surf. Sci.\/} {\bf 157} 43--60

\bibitem{grant84}
Grant R~B and Lambert R~M 1984 {\em Surf. Sci.\/} {\bf 146} 256--68

\bibitem{schmeisser85}
Schmeisser D and Jacobi K 1985 {\em Surf. Sci.\/} {\bf 156} 911--9

\bibitem{bartolucci}
Bartolucci F, Franchy R, Barnard J~C and Palmer R~E 1998 {\em Phys. Rev.
  Lett.\/} {\bf 80} 5224--5227

\bibitem{gravil96}
Gravil P~A, Bird D~M and White J~A 1996 {\em Phys. Rev. Lett.\/} {\bf 77}
  3933--6

\bibitem{monturet10}
Monturet S, Alducin M and Lorente N 2010 {\em Phys. Rev. B\/} {\bf 82} 085447

\bibitem{buatier97}
de~Mongeot F~B, Rocca M, Cupolillo A, Valbusa U, Kreuzer H~J and Payne S~H 1997
  {\em J. Chem. Phys.\/} {\bf 106} 711--8

\bibitem{carley90}
Carley A~F, Davies P~R, Roberts M~W and Thomas K~K 1990 {\em Surf. Sci.\/} {\bf
  238} L467--72

\bibitem{mongeot95}
de~Mongeot F~B, Valbusa U and Rocca M 1995 {\em Surf. Sci.\/} {\bf 339} 291--6

\bibitem{raukema95}
Raukema A, Dirksen R~J and Kleyn A~W 1995 {\em J. Chem. Phys.\/} {\bf 103}
  6217--31

\bibitem{raukema95prl}
Raukema A and Kleyn A~W 1995 {\em Phys. Rev. Lett.\/} {\bf 74} 4333--6

\bibitem{martin-gondre11}
Martin-Gondre L, Alducin M, Bocan G~A, Mui{\~n}o R~D and Juaristi J~I 2012
  Competition between electron and phonon excitations in the scattering of
  nitrogen off metal surfaces to be published

\bibitem{haberland85}
Haberland H, Buck U and Tolle M 1985 {\em Rev. Sci. Instrum.\/} {\bf 56}
  1712--6

\bibitem{raukema96diss}
Raukema A 1996 {\em Dynamics of Chemisorption\/} {Ph.D.} thesis University of
  Amsterdam

\end{thebibliography}

\providecommand{\newblock}{}

\end{document}